\input{aipcheck}

\documentclass[
    ,final            
   ,numberedheadings 
  ]
  {aipproc}

\layoutstyle{6x9}
\usepackage{upgreek}
\usepackage{latexsym}
\usepackage{amsmath,amsthm}
\usepackage{amssymb}
\usepackage{graphics}
\usepackage{marvosym}
\usepackage{color}
\newtheorem*{problem*}{Problem}
\newtheorem*{theorem*}{Theorem}
\newtheorem*{proposition*}{Proposition}

\newcommand{\nabb}{\mbox{$\nabla \mkern-13mu /$\,}}

\begin{document}

\title{The wave equation on axisymmetric stationary black hole backgrounds}


\author{Mihalis Dafermos}{
  address={University of Cambridge,
Department of Pure Mathematics and Mathematical Statistics,
Wilberforce Road, Cambridge CB3 0WB United Kingdom}
}

\begin{abstract}
Understanding the behaviour of linear waves on black hole backgrounds is a 
central problem in general relativity, intimately connected with the nonlinear stability of 
the black hole spacetimes themselves as solutions to the Einstein equations--a major 
open question in the subject. Nonetheless, it is only very recently that even the most 
basic boundedness and quantitative decay properties of linear waves have been proven 
in a suitably general class of black hole exterior spacetimes. This talk will review our 
current mathematical understanding of waves on black hole backgrounds, beginning with 
the classical boundedness theorem of Kay and Wald on exactly
Schwarzschild exteriors and ending 
with very recent boundedness and decay theorems (proven in collaboration with Igor Rodnianski)
on a wider class of spacetimes.
This class of spacetimes includes in particular slowly rotating Kerr spacetimes,
but in the case of the boundedness theorem is in fact much larger, encompassing
general axisymmetric stationary spacetimes whose geometry is sufficiently close
to Schwarzschild and whose Killing fields span the null generator
of the horizon.
\end{abstract}

\maketitle


\section{The problem}
Let $(\mathcal{M},g)$ be a black hole spacetime,
for instance Schwarzschild or Kerr\footnote{We refer the reader to standard texts~\cite{he:lssst, wald}
for a discussion of these spacetimes.}, but more generally, a spacetime whose geometry is ``near'' one
of the above. We will understand the meaning of ``near'' further down, so the reader may
for now wish to fix $(\mathcal{M},g)$ as precisely Schwarzschild or Kerr. 

Let $\Sigma$ denote an arbitrary Cauchy surface\footnote{For the purpose of this talk,
all spacetimes are globally hyperbolic. In particular, by the term ``Kerr spacetime''
we mean the globally hyperbolic subset of maximally extended Kerr consisting of the
development of a partial Cauchy
hypersurface with two asymptotically flat ends.}
for $(\mathcal{M},g)$.  It is known that for suitably regular initial data $\upPsi,\upPsi'$ prescribed
on $\Sigma$
for the wave equation
\begin{equation}
\label{thewaveequation}
\Box_g\psi=0,
\end{equation}
there exists a unique solution $\psi$  defined
globally on $\mathcal{M}$.

The problem of interest here is:
\begin{problem*}
Understand the quantitative boundedness and decay properties of $\psi$
in the closure $\mathcal{D}$ of the domain of outer communications of $(\mathcal{M},g)$.
\end{problem*}

Below is a  Penrose diagram indicating the region of interest in the case of Schwarzschild.
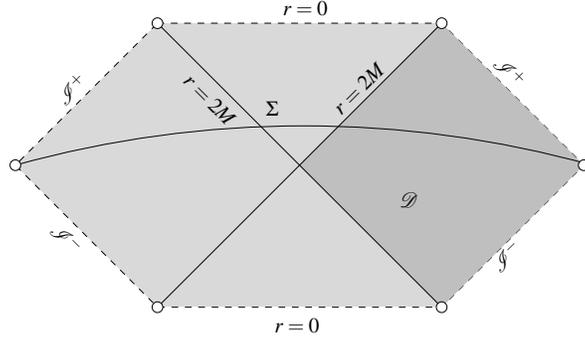
\[
\begin{picture}(0,0)%
\includegraphics{wCauchy.pstex}%
\end{picture}%
\setlength{\unitlength}{2763sp}%
\begingroup\makeatletter\ifx\SetFigFont\undefined%
\gdef\SetFigFont#1#2#3#4#5{%
  \reset@font\fontsize{#1}{#2pt}%
  \fontfamily{#3}\fontseries{#4}\fontshape{#5}%
  \selectfont}%
\fi\endgroup%
\begin{picture}(5239,3031)(1895,-7310)
\put(4201,-5311){\makebox(0,0)[lb]{\smash{{\SetFigFont{8}{9.6}{\rmdefault}{\mddefault}{\updefault}{\color[rgb]{0,0,0}$\Sigma$}%
}}}}
\put(3451,-5011){\rotatebox{315.0}{\makebox(0,0)[lb]{\smash{{\SetFigFont{8}{9.6}{\rmdefault}{\mddefault}{\updefault}{\color[rgb]{0,0,0}$r=2M$}%
}}}}}
\put(4876,-5311){\rotatebox{45.0}{\makebox(0,0)[lb]{\smash{{\SetFigFont{8}{9.6}{\rmdefault}{\mddefault}{\updefault}{\color[rgb]{0,0,0}$r=2M$}%
}}}}}
\put(2401,-5236){\rotatebox{45.0}{\makebox(0,0)[lb]{\smash{{\SetFigFont{8}{9.6}{\rmdefault}{\mddefault}{\updefault}{\color[rgb]{0,0,0}$\mathcal{I}^+$}%
}}}}}
\put(6226,-4861){\rotatebox{315.0}{\makebox(0,0)[lb]{\smash{{\SetFigFont{8}{9.6}{\rmdefault}{\mddefault}{\updefault}{\color[rgb]{0,0,0}$\mathcal{I}^+$}%
}}}}}
\put(6301,-6736){\rotatebox{45.0}{\makebox(0,0)[lb]{\smash{{\SetFigFont{8}{9.6}{\rmdefault}{\mddefault}{\updefault}{\color[rgb]{0,0,0}$\mathcal{I}^-$}%
}}}}}
\put(2251,-6361){\rotatebox{315.0}{\makebox(0,0)[lb]{\smash{{\SetFigFont{8}{9.6}{\rmdefault}{\mddefault}{\updefault}{\color[rgb]{0,0,0}$\mathcal{I}^-$}%
}}}}}
\put(4351,-4411){\makebox(0,0)[lb]{\smash{{\SetFigFont{8}{9.6}{\rmdefault}{\mddefault}{\updefault}{\color[rgb]{0,0,0}$r=0$}%
}}}}
\put(4276,-7261){\makebox(0,0)[lb]{\smash{{\SetFigFont{8}{9.6}{\rmdefault}{\mddefault}{\updefault}{\color[rgb]{0,0,0}$r=0$}%
}}}}
\put(5401,-6136){\makebox(0,0)[lb]{\smash{{\SetFigFont{8}{9.6}{\rmdefault}{\mddefault}{\updefault}{\color[rgb]{0,0,0}$\mathcal{D}$}%
}}}}
\end{picture}%

\]
We restrict in fact to 
\[
\mathcal{D}=\overline{J^+(\mathcal{I}^-_A)\cap J^-(\mathcal{I}^+_A)}
\]
where the closure refers to the topology of $\mathcal{M}$,
and where $\mathcal{I}^\pm_A$ denote a pair 
of connected components of $\mathcal{I}^\pm$, respectively,  with a 
common limit point.\footnote{Without loss of generality, we can restrict to this set
as opposed to $\overline{J^+(\mathcal{I}^-)\cap J^-(\mathcal{I}^+)}$. Astrophysical
black holes, of course, have only one asymptotically flat end.} We are thus interested in understanding
the behaviour of $\psi$ up to and including the event horizon $\mathcal{H}=\{r=2M\}$.
``Quantitative'' in the statement of our problem means we want to estimate
the size of $\psi$ in $\mathcal{D}$ in terms of quantities depending only on a 
suitable norm of the data
$\upPsi$, $\upPsi'$ on $\Sigma$.

The above problem is one of the most basic questions to pose about black hole spacetimes,
and is in fact intimately related to the non-linear stability problem of
the Kerr family as a family of solutions of the Einstein equations (see Section~\ref{last}).
Not surprisingly then, the problem has been the object of much study in general relativity, 
beginning with the work of Regge and Wheeler~\cite{RW}. Nonetheless,
until the past 5 years, the only result known for general solutions of
the Cauchy problem--i.e.~solutions not restricted by symmetry assumptions or
support assumptions--was the uniform boundedness of $\psi$ in $\mathcal{D}$,
in the very special case of Schwarzschild. 
This celebrated theorem of Kay and Wald is reviewed in Section~\ref{KWT}.

The rest of the talk will
then  review the recent progress in this area, which has now allowed
for a satisfactory answer to our motivating problem, not only for Schwarzschild itself,
but for spacetimes $(\mathcal{M},g)$ suitably
``near'' Schwarzschild, \emph{including the important Kerr and Kerr-Newman families (for small
parameters $a$, $Q$)}.
The main elements  central to our understanding of the problem can be summarised
by the following:
\begin{enumerate}
\item
A new, more robust proof of Kay and Wald's theorem making use of the red-shift effect,
ensuring good control at the horizon. (See Sections~\ref{stable?}--\ref{stronger}). The proof
turns out to be stable to a large class of
perturbations of the Schwarzschild metric, not however to Kerr! (See Section~\ref{perturb?}.)
\item
A proof of quantitative decay bounds for solutions of $(\ref{thewaveequation})$
on Schwarzschild. The main difficulties are (i) understanding
and quantifying the phenomenon of trapping
associated with the photon sphere, (ii) finding the analogue
for Schwarzschild of the conformal energy current
used to prove energy decay in Minkowski space and (iii) relating this 
to the red-shift effect, recovering decay near the horizon. (See Section~\ref{DeconS}.)
\item
The discovery that superradiant frequencies are not trapped for axisymmetric
stationary spacetimes sufficiently near Schwarzschild, allowing for a boundedness
theorem for all such spacetimes without a detailed understanding of trapping.
 This class of spacetimes includes Kerr and Kerr-Newman
for $a\ll M$, $Q\ll M$, but is in fact much more general. (See Section~\ref{Ubound}.) 
\item
Quantifying the trapping phenomenon on Kerr itself by frequency-localised\footnote{Defined
naturally relative to the geometric symmetries and hidden symmetries of Kerr} versions
of the virial identities used in Schwarzschild.
In view of the robustness of the other aspects of the decay proof on Schwarzschild, this
yields a proof of decay for solutions to $(\ref{thewaveequation})$ on Kerr
for $a\ll M$. (See Section~\ref{DecKer}.)
\end{enumerate}
The above serves also as an outline for the bulk of the talk. Let us emphasize that
the results outlined here do not close the book on this subject. 
What is the situation for higher spin? What are the least amount of assumptions
on the geometry which yield quantitative decay? What happens when the condition
$a\ll M$ is relaxed?
What is the relation with the non-linear stability of the background solutions themselves?
We end with remarks about future directions in Section~\ref{last}.

\section{Uniform boundedness on Schwarzschild}

\subsection{The Kay--Wald theorem}
\label{KWT}
The first definitive theorem in the direction of our motivating problem is the following
celebrated uniform boundedness
result for solutions of $(\ref{thewaveequation})$ on Schwarzschild exteriors.
\begin{theorem*}
(Kay--Wald~\cite{kw:lss}, 1987) Let $(\mathcal{M},g_M)$ be Schwarzschild with parameter
$M>0$, $\mathcal{D}$
as above the closure of its domain of outer communications, $\Sigma$ a Cauchy
surface for $\mathcal{M}$ and $\psi$ the unique solution of the wave equation $(\ref{thewaveequation})$ on
$\mathcal{M}$ with sufficiently regular initial data $\upPsi, \upPsi'$ on $\Sigma$,
decaying appropriately near spatial infinity $i^0$. Then there exists a $D$ 
depending only on the data such that 
\[
|\psi|\le D
\]
holds in $\mathcal{D}$.
\end{theorem*}

Before turning to the main conceptual difficulty of the proof of the above
theorem, let us make some general remarks.
The proofs of all theorems to be discussed in this talk
use ``energy type estimates'' to control square integral
quantities of $\psi$ and its derivatives; pointwise bounds are retrieved
at the last stage from these energy integrals and a Sobolev inequality.\footnote{The centrality of energy bounds in the study of the wave equation
arises from the fact that estimates of square integral quantities are the only 
estimates for solutions $\psi$ of $(\ref{thewaveequation})$ (in more than one spatial 
dimension) which do not lose derivatives.}

\subsection{Energy currents and vector fields}
Energy estimates for $(\ref{thewaveequation})$ have a very geometric origin 
which 
is intimately related to its Lagrangian structure. Let us briefly explain.

\subsubsection{Energy currents constructed from vector field multipliers}
Associated to the Lagrangian for $(\ref{thewaveequation})$ is the so called
energy-momentum tensor
\[
T_{\mu\nu}(\psi)=\partial_\mu\psi\partial_\nu\psi-\frac12g_{\mu\nu}g^{\alpha\beta}\partial_\alpha\psi
\partial_\beta\psi.
\]
For a solution $\psi$ to $(\ref{thewaveequation})$, $T_{\mu\nu}$ is divergence-free, i.e.
\begin{equation}
\label{divfree}
\nabla^\mu T_{\mu\nu}=0.
\end{equation}

Given any vector field $V$, we may associate to it two currents:
\[
J_\mu^V(\psi)\doteq T_{\mu\nu}(\psi) V^\nu, \qquad
K^V(\psi) \doteq 2\pi^{\mu\nu}_VT_{\mu\nu}(\psi),
\]
where $\pi^{\mu\nu}_V \doteq \frac12 V^{\mu;\nu}$ is the so called deformation tensor 
of $V$.
The relation $(\ref{divfree})$ yields
\[
K^V(\psi)=\nabla^\mu J_\mu(\psi)
\]
for solutions $\psi$ of $(\ref{thewaveequation})$. Thus, by the divergence theorem
If  $\Sigma_1$ and $\Sigma_2$ are homologous hypersurfaces bounding a spacetime
region $\mathcal{B}$, we have
\begin{equation}
\label{divthe}
\int_{\Sigma_2} J^V_\mu (\psi)n^\mu_{\Sigma_2}
+\int_{\mathcal{B}} K^V(\psi) =
\int_{\Sigma_1} J^V_\mu(\psi) n^\mu_{\Sigma_1}.
\end{equation}

When $V$ is timelike and $\Sigma_i$ is spacelike, then
$J^V_\mu (\psi)n^\mu_{\Sigma_2}\ge 0$, and in fact controls the spacetime
gradient of $\psi$.
If $V$ is in addition Killing, then $K^V=0$, and $(\ref{divthe})$ would provide
an estimate for the solution on $\Sigma_2$ from knowledge of the solution
on $\Sigma_1$ (``data''). Even when $V$ is not Killing, $K^V$ can sometimes
be treated as an error term.\footnote{For instance, 
the well-posedness for $(\ref{thewaveequation})$
can be proven by using $(\ref{divthe})$ for an arbitrary timelike $V$.}

One can also turn the identity $(\ref{divthe})$ on its head, and think about it as a way
to estimate 
\[
\int_{\mathcal{B}} K^V
\]
from the boundary terms. (Think about the classical virial theorem\ldots)
This is particularly useful when the boundary terms are controlled by a controlled
energy say, and when $K^V(\psi)\ge 0$ and controls derivatives of $\psi$.

Both uses of $(\ref{divthe})$ will arise in what follows.

\subsubsection{Vector fields as commutators and higher order currents}
In order to obtain pointwise bounds via energy control, one must consider ``higher order
energies''. 

Let us first consider the case where $W$ is a vector field in the Lie algrebra of
isometries of $g$. Then if $\psi$ satisfies $(\ref{thewaveequation})$, then so does
$W\psi$. More generally, if $W_1,\ldots W_k$ are in the Lie algebra, and $\psi$
satisfies $(\ref{thewaveequation})$, then so does $W_1\cdots W_k\psi$. Given a multiplier
vector field $V$, we may
thus consider the $k+1$'th order currents $J^{V}_\mu(W_1\cdots W_k\psi)$, and
$K^V(W_1\cdots W_k\psi)$. Again, $\nabla^\mu J^V_\mu=K^V$, and $(\ref{divthe})$
allows for proving higher-oder energy estimates.

If $W_1,\ldots W_k$ are not in the Lie algebra, one obtains an identity
\[
\nabla^\mu J^V_\mu(W_1\cdots W_k\psi) = K^V
(W_1\cdots W_k\psi)+ V^\mu\partial_\mu (W_1\cdots W_k\psi) F^{W_1,\dots W_k}
\]
where $K^V$, $J^V_\mu$ are defined as before,  and $F^{W_1,\dots W_k}$ 
is a current of order less than or equal to $k+1$. 
The above identity upon integration again allows for estimation of the higher
order energy $J^V_\mu(W_1\cdots W_k\psi)$.
For  a fundamental application of considering Lorentz boosts as commutators for proving decay for solutions of $(\ref{thewaveequation})$ on Minkowski space, see~\cite{muchT}. 

For a more general discussion of the origin of these identities for general Lagrangian
theories and their
relation to hyperbolicity, see the beautiful discussion in~\cite{book2}.

\subsection{The Kay--Wald proof}
Let us turn now to the proof of the Kay--Wald theorem, so as to see the main difficulty.
In Kay and Wald's proof, the only vector field used as a multiplier is 
\[
T=\frac\partial{\partial t},
\]
where $t$ is a Schwarzschild coordinate (in which the metric takes
the Schwarzschild form $-(1-2M/r)dt^2+(1-2M/r)^{-1} dr^2+r^2d\gamma_{\mathbb S^2}$
in the interior of $\mathcal{D}$). Recall that the vector field $T$ extends to a Killing field
on all $\mathcal{M}$, is
timelike in the interior
of $\mathcal{D}$, and null on its boundary $\mathcal{H}^+\cup\mathcal{H}^-$, vanishing
on the sphere of bifurcation $\mathcal{H}^+\cap\mathcal{H}^-$.

Without loss of generality, we may assume that our Cauchy surface $\Sigma$ intersected
with $\mathcal{D}$ is as depicted below by $\Sigma_0$:
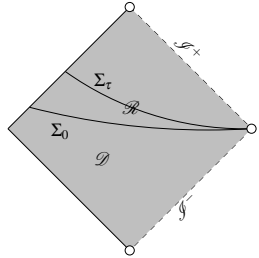
\[
\begin{picture}(0,0)%
\includegraphics{newnewfol.pstex}%
\end{picture}%
\setlength{\unitlength}{2368sp}%
\begingroup\makeatletter\ifx\SetFigFont\undefined%
\gdef\SetFigFont#1#2#3#4#5{%
  \reset@font\fontsize{#1}{#2pt}%
  \fontfamily{#3}\fontseries{#4}\fontshape{#5}%
  \selectfont}%
\fi\endgroup%
\begin{picture}(2618,2660)(4489,-7091)
\put(6226,-4861){\rotatebox{315.0}{\makebox(0,0)[lb]{\smash{{\SetFigFont{7}{8.4}{\rmdefault}{\mddefault}{\updefault}{\color[rgb]{0,0,0}$\mathcal{I}^+$}%
}}}}}
\put(6301,-6736){\rotatebox{45.0}{\makebox(0,0)[lb]{\smash{{\SetFigFont{7}{8.4}{\rmdefault}{\mddefault}{\updefault}{\color[rgb]{0,0,0}$\mathcal{I}^-$}%
}}}}}
\put(4951,-5836){\makebox(0,0)[lb]{\smash{{\SetFigFont{7}{8.4}{\rmdefault}{\mddefault}{\updefault}{\color[rgb]{0,0,0}$\Sigma_0$}%
}}}}
\put(5401,-6136){\makebox(0,0)[lb]{\smash{{\SetFigFont{7}{8.4}{\rmdefault}{\mddefault}{\updefault}{\color[rgb]{0,0,0}$\mathcal{D}$}%
}}}}
\put(5401,-5311){\makebox(0,0)[lb]{\smash{{\SetFigFont{7}{8.4}{\rmdefault}{\mddefault}{\updefault}{\color[rgb]{0,0,0}$\Sigma_\tau$}%
}}}}
\put(5701,-5611){\makebox(0,0)[lb]{\smash{{\SetFigFont{7}{8.4}{\rmdefault}{\mddefault}{\updefault}{\color[rgb]{0,0,0}$\mathcal{R}$}%
}}}}
\end{picture}%

\]
We may define a regular coordinate system $(r, t^*)$ in
\[
\mathcal{R}\doteq \mathcal{D}\cap J^+(\Sigma_0)
\]
such that $\Sigma_0$ corresponds to $t^*=0$ and $T$ still corresponds to $\partial_{t^*}$.
We may define then $\Sigma_\tau=\{t^*=\tau\}$. Note that $\Sigma_\tau = \varphi_\tau(\Sigma_0)$,
where $\varphi_s$ denotes the one-parameter group of diffeomorphisms generated by $T$.

Let us apply the energy identity of $J^T$ in the region bounded by $\Sigma_0$,
$\Sigma_\tau$, and the corresponding piece of $\mathcal{H}$.
On $\Sigma_\tau$, one has
\[
J^T_\mu(\psi)n^\mu_{\Sigma_\tau} \sim \left((1-2M/r)(\partial_r\psi)^2 +(\partial_{t^*}\psi)^2
+|\nabb\psi|^2\right)
\]
where $|\cdot|$, $\nabb$ denote here the induced norm and
connection in the ${\rm SO}(3)$ group orbits.
Note the degeneration of the $\partial_r$ derivative at $\Sigma_\tau\cap\mathcal{H}$.
This arises because $T$ becomes null on $\mathcal{H}$.

Since the flux through the horizon is nonnegative
\begin{equation}
\label{afou}
J^T_\mu(\psi)n^\mu_{\mathcal{H}} \ge 0
\end{equation}
we have
\begin{align}
\label{schest}
\nonumber
\int_{\Sigma_\tau}& \left((1-2M/r)(\partial_r\psi)^2 +(\partial_{t^*}\psi)^2
+|\nabb\psi|^2\right)\\
\le 
&B \int_{\Sigma_0}\left((1-2M/r)(\partial_r\psi)^2 +(\partial_{t^*}\psi)^2
+|\nabb\psi|^2\right).
\end{align}

Commuting with $T$, i.e.~considering the current $J^T(T\psi)$ and $J^T(TT\psi)$,
one obtains $(\ref{schest})$ with $\psi$ replaced by $T\psi$ and $TT\psi$. 
An elliptic estimate and the Sobolev inequality suffices to show that if
$\lim_{x\to i^0}\upPsi=0$, then 
\[
(1-2M/r) \psi^2 \le  \int_{\Sigma_0} (J^T_\mu(\psi)+J^T_\mu(T\psi)+J^T_\mu(T\psi))n^\mu_{\Sigma_0} 
\]
in $\mathcal{R}$.

The above argument is in fact completely standard and yields the statement of the theorem
but where $\psi$ is replaced by $\sqrt{1-\frac{2M}r}\psi$. Thus, it provides no information about
the behaviour of $\psi$ along 
$\mathcal{H}$ where $r=2M$!
\emph{Understanding the behaviour up to and including the horizon is thus
the only real difficulty of this problem in the Schwarzschild case.}

This difficulty is overcome by Kay and Wald by the following argument:
One first notices that if there exists a $\tilde\psi$ satisfying 
$(\ref{thewaveequation})$ such that $T\tilde\psi=\psi$, then one can estimate
$\psi$ by a suitable Sobolev inequality from the energies of $\tilde\psi$ on
$\{t=c\}$, providing one also commute with all angular momentum operators
$\Omega_i$. Such a $\tilde\psi$ can be constructed if it is assumed that
$\psi$ is not supported in a neighborhood of $\mathcal{H}^+\cap\mathcal{H}^-$, 
by inverting an elliptic operator on $\Sigma_0$.
(This was in fact an earlier observation of Wald.)
More generally, $\tilde\psi$ can be constructed if $\psi$ decays
suitably to $0$ at $\mathcal{H}^+\cap
\mathcal{H}^-$. But what to do in the general case where $\psi$ is not
assumed to have special behaviour at $\mathcal{H}^+\cap\mathcal{H}^-$?

Here comes the second clever observation: Since one is only interested
in the behaviour in $\mathcal{R}$, one can replace $\psi$ by a solution $\hat\psi$
of $(\ref{thewaveequation})$ such that $\psi=\hat\psi$ in $\mathcal{R}$. Using the domain
of dependence property for the wave equation, the discrete symmetry of (extended)
Schwarzschild interchanging the two ends, and a preservation of symmetry
argument, one can construct a $\hat\psi$ with the 
desired behaviour at $\mathcal{H}^+\cap\mathcal{H}^-$. One can then construct
$\tilde{\hat\psi}$ and continue as before.

See~\cite{kw:lss} or~\cite{notes} for more details.

\subsection{A stable proof?}
\label{stable?}
The above proof is remarkable, but fragile! It requires (i) the staticity property to construct
$\tilde\psi$, (ii) the spherical symmetry of Schwarzschild as one must commute 
$(\ref{thewaveequation})$ with $\Omega_i$, $i=1,\ldots 3$, and, finally, even (iii) the
discrete symmetry of Schwarzschild. Is it really the case that a result so fundamental
as boundedness must depend on all this special structure of Schwarzschild?

The difficulties of the proof arise because the set of multipliers and commutators
are restricted to the Killing fields $T$, and $\Omega_i$. There is another important
physical property of Schwarzschild which is not apparent from these alone: We discuss
 this in the next section.

\subsection{The redshift effect}
The \emph{red-shift effect} is one of the most
celebrated aspects of black holes. It is classically described as follows:
 Suppose two observers, $A$ and $B$ are such that
$A$ crosses the event horizon and $B$ does not. If $A$ emits a signal at constant
frequency as he measures it, then the frequency at which it is received by
$B$ is ``shifted to the red''. 
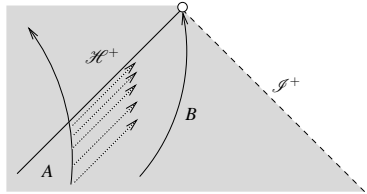
\[
\begin{picture}(0,0)%
\includegraphics{redshift.pstex}%
\end{picture}%
\setlength{\unitlength}{2368sp}%
\begingroup\makeatletter\ifx\SetFigFont\undefined%
\gdef\SetFigFont#1#2#3#4#5{%
  \reset@font\fontsize{#1}{#2pt}%
  \fontfamily{#3}\fontseries{#4}\fontshape{#5}%
  \selectfont}%
\fi\endgroup%
\begin{picture}(3763,2002)(4950,-6673)
\put(6826,-5911){\makebox(0,0)[lb]{\smash{{\SetFigFont{7}{8.4}{\rmdefault}{\mddefault}{\updefault}{\color[rgb]{0,0,0}$B$}%
}}}}
\put(5776,-5311){\makebox(0,0)[lb]{\smash{{\SetFigFont{7}{8.4}{\rmdefault}{\mddefault}{\updefault}{\color[rgb]{0,0,0}$\mathcal{H}^+$}%
}}}}
\put(7726,-5611){\makebox(0,0)[lb]{\smash{{\SetFigFont{7}{8.4}{\rmdefault}{\mddefault}{\updefault}{\color[rgb]{0,0,0}$\mathcal{I}^+$}%
}}}}
\put(5326,-6511){\makebox(0,0)[lb]{\smash{{\SetFigFont{7}{8.4}{\rmdefault}{\mddefault}{\updefault}{\color[rgb]{0,0,0}$A$}%
}}}}
\end{picture}%

\]
The consequences of this for the appearance of a  collapsing star  to far-away observers
were first explored in the seminal paper of Oppenheimer-Snyder~\cite{os}.

The red-shift effect as described above is a global one, and essentially depends only
on the fact that the proper time of $B$ is infinite whereas the proper time of $A$ before crossing
$\mathcal{H}^+$ is finite.
In the case of the Schwarzschild black hole, 
there is a ``local'' version of this red-shift: If $B$ also crosses the event 
horizon but at advanced time later than $A$:
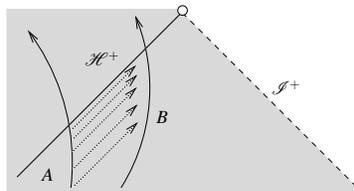
\[
\begin{picture}(0,0)%
\includegraphics{locredshift.pstex}%
\end{picture}%
\setlength{\unitlength}{2368sp}%
\begingroup\makeatletter\ifx\SetFigFont\undefined%
\gdef\SetFigFont#1#2#3#4#5{%
  \reset@font\fontsize{#1}{#2pt}%
  \fontfamily{#3}\fontseries{#4}\fontshape{#5}%
  \selectfont}%
\fi\endgroup%
\begin{picture}(3763,2002)(4950,-6673)
\put(5776,-5311){\makebox(0,0)[lb]{\smash{{\SetFigFont{7}{8.4}{\rmdefault}{\mddefault}{\updefault}{\color[rgb]{0,0,0}$\mathcal{H}^+$}%
}}}}
\put(7726,-5611){\makebox(0,0)[lb]{\smash{{\SetFigFont{7}{8.4}{\rmdefault}{\mddefault}{\updefault}{\color[rgb]{0,0,0}$\mathcal{I}^+$}%
}}}}
\put(5326,-6511){\makebox(0,0)[lb]{\smash{{\SetFigFont{7}{8.4}{\rmdefault}{\mddefault}{\updefault}{\color[rgb]{0,0,0}$A$}%
}}}}
\put(6526,-5911){\makebox(0,0)[lb]{\smash{{\SetFigFont{7}{8.4}{\rmdefault}{\mddefault}{\updefault}{\color[rgb]{0,0,0}$B$}%
}}}}
\end{picture}%

\]
 then the 
frequency at which $B$ receives at his horizon crossing time
is shifted to the red by a factor depending exponentially on the advanced time difference
of the crossing points of $A$ and $B$.

The exponential factor is determined by the so-called \emph{surface gravity},
a quantity that can in fact be defined for all so-called Killing horizons. This localised red-shift
effect depends only on the positivity of this quantity.

\subsection{The redshift as seen by vector fields}
\label{red}
It turns out that the local red-shift effect can be captured by positivity
properties in the energy identity of a suitably constructed
vector field multiplier applied both to $\psi$ alone and to $\psi$ commuted with
a suitably constructed vector field commutator.

\begin{proposition*}\cite{dr3, notes}
There exists a smooth vector field $N$, and two positive constants $0<b<B$
such that $N$ is timelike and $\varphi_t$-invariant
such that
\begin{equation}
\label{upologismos}
 b\, J^N_\mu(\psi ) N^\mu \le K^N (\psi)\le B\, J^N_\mu(\psi)  N^\mu,
\end{equation}
along $\mathcal{H}^+$,
for all solutions $\psi$ of $\Box_g\psi=0$.
\end{proposition*}

A vector-field commutator version can be seen by
\begin{proposition*}\cite{bounded, notes}
Under the assumptions of the above theorem, let $Y=N-T$, and extend $T$, $Y$ to 
a null frame $T, Y, E_1, E_2$ on $\mathcal{H}^+$.
If $\psi$ satisfies $\Box_g\psi=0$, then for all $k \ge 1$.
\begin{equation}
\label{comcomp}
\Box_g( Y^k\psi) = b_k Y^k\psi + \sum_{0\le |m|\le k,\, 0\le m_4<k }c_m E_1^{m_1}
E_2^{m_2}T^{m_3}Y^{m_4} \psi
\end{equation}
on $\mathcal{H}^+$, where $b_k>0$.
\end{proposition*}

These propositions apply in particular to the Schwarzschild metric, but in fact, their
domain of validity is much more general: They apply to any stationary black hole
with event horizon with positive surface gravity. See~\cite{notes}.

\subsection{A stronger boundedness theorem}
\label{stronger}
The above ``positive terms'', $K^N$ and $b_k Y^k\psi$ can be viewed as exponential 
damping terms in the energy identities with $N$ as a multiplier, 
and more generally, $N$ as a multiplier applied to $\psi$ commuted with $Y^k$.
Of course, these nice properties hold only near the horizon. Thus, to use these identities
one must apply these estimates \emph{in conjunction} with a statement giving
good control away from the horizon. In the case of Schwarzschild, this good control
follows from the first part of the argument described below, i.e.~from application of $T$ to $\psi$,
$T\psi$, $TT\psi$. This allows for a proof of the following stronger boundedness
statement.
\begin{theorem*}~\cite{notes} 
Let $(\mathcal{M},g_M)$ be Schwarzschild and $\Sigma_\tau$ as above.
Then there exists a constant $C$ depending
only on $M$, $\Sigma_0$
such that for all $\psi$ satisfying $\Box_g\psi=0$,
the following holds:
\[
|n_{\Sigma_\tau}\psi|_{L^2(\Sigma_\tau)}
+  |\nabla_{\Sigma_\tau}\psi|_{L^2(\Sigma_\tau)}\le C
(|n_{\Sigma}\psi|_{L^2(\Sigma)}
+|\nabla_{\Sigma}\psi|_{L^2(\Sigma)}).
\]

Moreover, for all $m\ge0$, the $m$'th order pointwise bounds
\[
\sum_{0\le m_1+m_2\le m}|\nabla_{\Sigma_\tau}^{(m_1)}n^{(m_2)}_{\Sigma_\tau}\psi|\le C\, {\bf Q}_m
\]
hold in $\mathcal{R}$,
where ${\bf Q}_m$ is an appropriate norm on initial data.
\end{theorem*}

The above theorem is stronger than the Kay and Wald statement in that it proves the uniform
boundedness of 
an ``energy'' which does not degenerate in local coordinates on the horizon.\footnote{That is
to say, the energy computed by a $\varphi_t$-invariant family of freely falling observers is
proven bounded.} Moreover, this boundedess is proven for arbitrary higher order energies,
leading to pointwise bounds for arbitrary derivatives, including transversal derivatives
to the horizon.
\emph{It is interesting to remark that the Kay--Wald argument cannot prove
the uniform boundedness of these transversal derivatives.}

\subsection{Perturbing the metric}
\label{perturb?}
The above proof now is much more robust.  In fact, 
it can be perturbed to nearby metrics as long as one retains $\mathcal{H}^+$ as
a null boundary and $T$ as Killing and causal:

\begin{theorem*}~\cite{notes}
Let $\mathcal{R}$, $T$ be as before,
and let $g$ be a metric on $\mathcal{R}$ 
sufficiently close to Schwarzschild such that
$T$ is Killing and causal on $\mathcal{R}$, and $\mathcal{H}^+$ is null with respect to $g$.
Then the statement of the previous theorem applies verbatim.
\end{theorem*}

In view of the remarks at the end of Section~\ref{red}, it follows that one may weaken 
the assumption ``$g$ sufficiently close to Schwarzschild'', replacing it with the assumption 
that the geometry is that of a black hole with positive surface gravity. This and the remaining
assumptions are then in particular satisfied by all the classical static electrovacuum 
black holes
(Reissner-Nordstr\"om-de Sitter, etc.) Moreover, 
one need not assume that $T$ is Killing, merely that $\pi^T_{\mu\nu}$ decays
appropriately in $\tau$. See~\cite{notes} for details.

\emph{What about Kerr?}
Unfortunately, for all $a\ne 0$, the stationary vector field $T$ 
of the Kerr metric $g_{M,a}$ 
is no longer causal in the interior 
of $\mathcal{D}$ and thus $g_{M,a}$ does not satisfy the assumptions of the above
Theorem!
In particular, $(\ref{afou})$ does not hold and we can thus no longer a priori infer the
uniform boundedness of 
\[
\int_{\Sigma_\tau} J^T_\mu (\psi)n^\mu_{\Sigma_\tau}
\]
from the energy identity of $J^T$.

The part of $\mathcal{D}$ where $T$ is spacelike is known as the \emph{ergoregion}, 
and the associated behaviour of waves is known as \emph{superradiance}.
The test-particle manifestation of this fact is the celebrated 
\emph{Penrose process}. See~\cite{wald} for a nice discussion.

The above suggests that it may be difficult to prove boundedness alone, and that of
necessity one must try to prove more than boundedness at the same time, i.e.~decay.

\section{Decay on Schwarzschild}
\label{DeconS}
Before contemplating discussing decay for solutions to $(\ref{thewaveequation})$ 
on Kerr, we must first understand how such results can be proven on Schwarzschild. 
Some non-quantitative results, i.e.~decay without a rate~\cite{twainy},
scattering and asymptotic completeness statements~\cite{bachelot}, have been known for
some time. In view of our motivation in the problem of non-linear stability of the background
spacetime (see Section~\ref{last}), we are here interested exclusively
in quantitative statements: \emph{rates} of decay depending
only on the \emph{size} of initial data.

\subsection{The pointwise and energy decay theorem}
To talk about energy decay on Schwarzschild, one must 
introduce a different type of foliation.
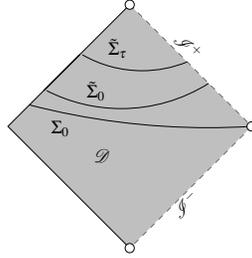
\[
\begin{picture}(0,0)%
\includegraphics{othernewfol.pstex}%
\end{picture}%
\setlength{\unitlength}{2368sp}%
\begingroup\makeatletter\ifx\SetFigFont\undefined%
\gdef\SetFigFont#1#2#3#4#5{%
  \reset@font\fontsize{#1}{#2pt}%
  \fontfamily{#3}\fontseries{#4}\fontshape{#5}%
  \selectfont}%
\fi\endgroup%
\begin{picture}(2618,2660)(4489,-7091)
\put(6301,-6736){\rotatebox{45.0}{\makebox(0,0)[lb]{\smash{{\SetFigFont{7}{8.4}{\rmdefault}{\mddefault}{\updefault}{\color[rgb]{0,0,0}$\mathcal{I}^-$}%
}}}}}
\put(6226,-4861){\rotatebox{315.0}{\makebox(0,0)[lb]{\smash{{\SetFigFont{7}{8.4}{\rmdefault}{\mddefault}{\updefault}{\color[rgb]{0,0,0}$\mathcal{I}^+$}%
}}}}}
\put(5326,-5461){\makebox(0,0)[lb]{\smash{{\SetFigFont{7}{8.4}{\rmdefault}{\mddefault}{\updefault}{\color[rgb]{0,0,0}$\tilde\Sigma_0$}%
}}}}
\put(4951,-5836){\makebox(0,0)[lb]{\smash{{\SetFigFont{7}{8.4}{\rmdefault}{\mddefault}{\updefault}{\color[rgb]{0,0,0}$\Sigma_0$}%
}}}}
\put(5551,-5011){\makebox(0,0)[lb]{\smash{{\SetFigFont{7}{8.4}{\rmdefault}{\mddefault}{\updefault}{\color[rgb]{0,0,0}$\tilde\Sigma_\tau$}%
}}}}
\put(5401,-6136){\makebox(0,0)[lb]{\smash{{\SetFigFont{7}{8.4}{\rmdefault}{\mddefault}{\updefault}{\color[rgb]{0,0,0}$\mathcal{D}$}%
}}}}
\end{picture}%

\]
Let $\Sigma$ be the Cauchy hypersurface as before (say coinciding with a surface
 $\{t=c\}$ for
all sufficiently large
$r$), and 
let $\tilde\Sigma$ now be a hypersurface with $\tilde\Sigma\subset J^+(\Sigma)$
such that $\tilde\Sigma\cap \mathcal{H}^+\ne\emptyset$, and $\tilde\Sigma$ meets $\mathcal{I}^+$
appropriately, and define $\tilde\Sigma_0=\tilde\Sigma\cap\mathcal{D}$,
$\tilde\Sigma_\tau=\varphi_\tau(\tilde
\Sigma_0)$ for $\tau\ge 1$.

\begin{theorem*}~\cite{dr3}
Let $(\mathcal{M},g_M)$ be Schwarzschild with parameter $M$, let $\Sigma$, $\tilde\Sigma$,
$\mathcal{D}$ as above, and let $\Omega_i$ denote the angular momentum operators.
Then there exists a constant $C$ depending
only on $M$, $\Sigma$ and $\tilde\Sigma$ such that for all $\psi$ satisfying $\Box_g\psi=0$,
the following holds:
\begin{align}
\label{withlo}
\nonumber
|n_{\tilde\Sigma_\tau}\psi|_{L^2(\tilde\Sigma_\tau)}
&+  |\nabla_{\tilde\Sigma_\tau}\psi|_{L^2(\tilde\Sigma_\tau)}\\
&\le C\tau^{-1}\sum_{|m|\le 3}
(r|n_{\Sigma}\Omega^{m}\psi|_{L^2(\Sigma)}
+r|\nabla_{\Sigma}\Omega^{m}\psi|_{L^2(\Sigma)}).
\end{align}
Moreover, the pointwise decay rates
\begin{equation}
\label{pwise}
|\sqrt{r} \psi|\le C\, {\bf Q}\, \tau^{-1}, \qquad |r\psi|\le C{\bf Q}\, \tau^{-1/2}
\end{equation}
hold,
where ${\bf Q}$ is an appropriate norm on initial data.
\end{theorem*}

One can in fact show decay for non-degenerate
energies of arbitrary order, and pointwise decay for arbitrary derivatives of $\psi$, including
derivatives transverse to the horizon. See~\cite{notes}.

An independent proof of similar decay rates away from the horizon but weaker decay
rates along the horizon was given by Blue and Sterbenz~\cite{BlueSter}.

\subsection{Trapping}
Before turning to the proof of the above theorem, let us
point out a central feature of its statement:
The energy decay estimate $(\ref{withlo})$
``loses'' derivatives, that is to say, one needs control of
more derivatives initially on $\Sigma$ to estimate the energy later on $\tilde\Sigma_\tau$.
This is an essential aspect of the problem and has to do with {\bf trapping}, i.e.~the fact that
there are null geodesics neither crossing the event horizon nor approaching null infinity.
These in fact asymptote to the so-called \emph{photon sphere} at $r=3M$:
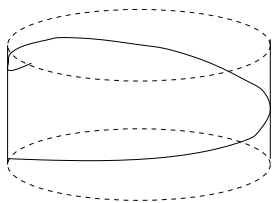
\[
\begin{picture}(0,0)%
\includegraphics{photon.pstex}%
\end{picture}%
\setlength{\unitlength}{1973sp}%
\begingroup\makeatletter\ifx\SetFigFont\undefined%
\gdef\SetFigFont#1#2#3#4#5{%
  \reset@font\fontsize{#1}{#2pt}%
  \fontfamily{#3}\fontseries{#4}\fontshape{#5}%
  \selectfont}%
\fi\endgroup%
\begin{picture}(3328,2414)(4937,-5468)
\end{picture}%

\]
which is itself spanned by null geodesics.

A rigorous study of the geometric optics approximation easily shows that
one can construct a sequence of solutions to $(\ref{thewaveequation})$
with fixed initial energy, such that the energy concentrates 
near such a trapped null geodesic for longer and longer time. \emph{This sequence
shows that an estimate of the form $(\ref{withlo})$ cannot hold without losing derivatives.}

\subsection{The vector fields}
The proof of decay uses multipliers constructed from $4$ different vector fields.
\subsubsection{The vector field $T$}
We have already discussed the use of this in the context of the Kay and Wald theorem.

\subsubsection{Trapping and the vector field $X$}
\label{Xconstr}
In the obstacle problem on Euclidean space, trapping is often ``captured'' by
the bulk term of the identity $(\ref{divthe})$ 
for multipliers corresponding to well chosen vector fields $X=f(r)\partial_r$. 
Soffer and collaborators in their pioneering~\cite{labasoffer, BlueSof0} 
were the first to pursue the programme of constructing such vector fields to capture
the trapping phenomenon in Schwarzschild.
The programme was first successfully completed in \cite{dr3} and~\cite{BlueSof}, but
using spherical harmonic decompositions. 
The multiplier to be discussed here, the first not to require such decompositions, was
constructed in~\cite{dr5}.

Le us recall first so-called Regge-Wheeler coordinates
$(r^*,t)$, where 
$r^*$ is defined by   
\begin{equation}
\label{rw}
r^*= r+2M\log (r-2M)-3M-2M\log M.
\end{equation}

The current ``capturing'' trapping
is actually a higher order current, involving also commutation, and takes the form
\begin{eqnarray}
\label{thecurrent}
\nonumber
J^{\bf X}_\mu(\psi) &=&
e J^N_\mu(\psi) + J^{X^a}_\mu(\psi)+ \sum_iJ^{X^b,w^b}_\mu(\Omega_i\psi)\\
&&\hbox{}-
\frac12\frac{r(f^b)'}{f^b(r-2M)}\left(\frac{r-2M}{r^2}-\frac{(r^*-\alpha-\alpha^{1/2})}{
\alpha^2+(r^*-\alpha-\alpha^{1/2})^2}\right) X^b_\mu \psi^2.
\end{eqnarray}
Here, $N$ is as in Section~\ref{red}, $X^a=f^a \partial_{r^*}$, $X^b=f^b\partial_{r^*}$,
the modified current $J^{X,w}$ is defined by
\begin{equation}
\label{warped}
J^{X,w}_\mu = X^\nu T_{\mu\nu} +\frac18w\partial_\mu(\psi^2)-\frac18(\partial_\mu w)\psi^2,
\end{equation}
and
\begin{equation}
\label{fdef}
f^a = -\frac{C_a}{\alpha r^2} + \frac{c_a}{r^{3}}, \qquad
f^b= \frac{1}{\alpha}\left(\tan^{-1}\frac{r^*-\alpha-\alpha^{1/2}}\alpha -\tan^{-1}(-1-\alpha^{-1/2})\right),
\end{equation}
\[
w^b= 
\frac{1}{8}\left((f^b)'+2\frac{r-2M}{r^2}f^b\right),
\]
and $e$, $C_a$, $c_a$, $\alpha$ are positive parameters which must be chosen accordingly.
With these choices, one can show (after some computation)
that the divergence $K^{\bf X}=\nabla^\mu J_\mu^{\bf X}$ 
controls in particular
\begin{equation}
\label{cip}
\int_{\mathbb S^2} K^{\bf X}(\psi)  \ge
b\chi  \int_{\mathbb S^2} J^N_\mu(\psi) n^\mu,
\end{equation}
where $\chi$ is non-vanishing but decays (polynomially) as $r\to\infty$, and the integration
is over any $SO(3)$ orbit.
Note that in view of the normalisation $(\ref{rw})$ of the $r^*$ coordinate, 
$X^b=0$ precisely at $r=3M$. 
The left hand side of the inequality $(\ref{cip})$ controls also
second order derivatives which degenerate however at $r=3M$. We have dropped these
terms. It is actually useful for applications that the $J^{X^a}(\psi)$ part of the current
is not ``modified'' by a function $w^a$, and thus
$\psi$ itself does not occur in the boundary terms.
That is to say
\begin{equation}
\label{demek=}
|J_\mu^{\bf X}(\psi) n^\mu | \le B\left(J^N_\mu(\psi) n^\mu+ 
\sum_{i=1}^3J^N_\mu(\Omega_i\psi)n^\mu \right).
\end{equation}
On the event horizon $\mathcal{H}^+$, we have a better one-sided bound
\begin{equation}
\label{better1}
-J_\mu^{\bf X}(\psi) n^\mu_{\mathcal{H}^+} \le B \left(J^T_\mu(\psi)n^\mu_{\mathcal{H}^+}+
\sum_{i=1}^3 J^T_\mu(\Omega_i\psi)n^\mu_{\mathcal{H}^+}\right).
\end{equation}
For details of the construction, see~\cite{dr5}.

In view of $(\ref{cip})$, $(\ref{demek=})$ and $(\ref{better1})$, together
with our previous boundedness theorem of Section~\ref{stronger}, 
one obtains in particular the estimate
\begin{equation}
\label{finalestimate}
\int_{J^+(\tilde\Sigma(\tau'))\cap\mathcal{D}} \chi J^N_\nu(\psi) n^\nu_{\tilde{\Sigma}}
      \le B \int_{\tilde\Sigma(\tau')}    \left(       J^N_\mu(\psi)  +         \sum_{i=1}^3 J^N_\mu
      (\Omega_i \psi)\right)
      n^\mu_{\tilde\Sigma_\tau},
\end{equation}
for some  nonvanishing $\varphi_t$-invariant function
$\chi$ which decays polynomially
as $r\to\infty$.
Such estimates are known as \emph{integrated decay}.

For a sketch of
yet another construction yielding an estimate $(\ref{finalestimate})$ which degenerates
however on $\mathcal{H}^+$, see~\cite{metcalfe}.\footnote{This degeneration can be
overcome by adding the energy identity of the current $N$ of Section~\ref{red}.}

\subsubsection{The vectorfield $Z$}
To turn this integrated decay into decay of energy as in the statement~$(\ref{withlo})$, 
one introduces a current
$J^{Z,w}$ (of the form $(\ref{warped})$) associated to a vector field $Z$ defined by
\begin{equation}
\label{Zdef}
u^2\partial_u +v^2\partial_v
\end{equation}
where $u=t-r^*$, $v=t+r^*$, and
\[
w= \frac{2tr^*(1-2M/r)}{r}.
\]
In the case of Minkowski space ($M=0$), the divergence $K^{Z,w}=0$, while
\begin{equation}
\label{touZ}
\int_{t=\tau} J^{Z,w}_\mu n^\mu \ge b\int_{t=\tau} u^2(\partial_u\psi)^2 +v^2(\partial_v\psi)^2 +
(u^2+v^2)|\nabb\psi|^2. 
\end{equation}
The identity $(\ref{divthe})$ yields the boundedness of the left hand side above,
and thus, in view of the weights on the right hand side of $(\ref{touZ})$, 
this yields decay of energy as in $(\ref{withlo})$.\footnote{Note that the
current  $J^{Z,w}_\mu$ is related to the conformal covariance properties of the wave equation on Minkowski space.}

In the case of Schwarzschild, a similar relation to $(\ref{touZ})$ holds (with an extra factor
of $(1-2M/r)$). But now the error term $K^{Z,w}\ne 0$, in fact the best one can estimate is
\begin{equation}
\label{otr}
-K^{Z,w} \ge B \, t J^N_\mu n^\mu
\end{equation}
in a region $[r_1,R_2]$ for some $R_2>r_1>2M$.

The error term on the right hand side of $(\ref{otr})$ at first seems problematic, but
it can in fact be absorbed by a simple iteration argument\footnote{using also the considerations 
of Section~\ref{Nfordecay} below}, 
given only the integrated decay estimate $(\ref{finalestimate})$. 
Thus one retrieves energy decay statements on Schwarzschild exactly analogous
to the case of Minkowski space, but now ``losing''
derivatives, in view of the use of $(\ref{finalestimate})$
to absorb the error term above.

Note that a related method of absorbing the error term on the right hand side of $(\ref{otr})$
was independently attained in the paper~\cite{BlueSter} referred to previously. 

\subsubsection{The vectorfield $N$}
\label{Nfordecay}
The above does not give proper control at the horizon. For this, one must return to
the vector field $N$ of Section~\ref{red}. It turns out that the calculation $(\ref{upologismos})$,
 in conjunction with the bounds obtained away from the horizon, allows one to extend the energy decay and pointwise decay results to the horizon.
For details, see~\cite{dr3, dr5} or~\cite{notes}.

\subsection{Commutation and Sobolev inequalities}
To achieve pointwise control $(\ref{pwise})$ from $(\ref{withlo})$, we commute with $\Omega_i$ and
apply Sobolev inequalities. See~\cite{dr3}.

\subsection{Price law tails?} 
In 1972, Price~\cite{rpr:ns} put forth heuristic arguments 
suggesting that, decomposing $\psi$ into spherical harmonics $\psi_\ell$,
 each $\psi_\ell$ should asymptotically
behave asymptotically like
\begin{equation}
\label{price}
\psi_\ell (r, t) \sim C_\ell t^{-(3+2\ell)} .
\end{equation}
Related statements have indeed been proven in the case $\ell=0$ (see~\cite{dr1}), 
but no statement
of the form $(\ref{price})$ has yet been shown for general $\ell$. 

Recall that the our interest in the linear theory is motivated by the desire to 
understand the non-linear stability problem (See Section~\ref{last}).
For this, a statement of the form
$(\ref{price})$ would be essentially useless:
The statement $(\ref{price})$, even if true, would be completely
non-quantitative, i.e.~it would not give a bound for $\psi_\ell$ at ``intermediate times''
in terms of the size of initial data.
In particular, the statement $(\ref{price})$ would not ``see'' the trapping
phenomenon and the associated loss of derivatives in the estimate $(\ref{withlo})$.

One faces this non-quantitative aspect immediately when one tries to sum
$(\ref{price})$ over $\ell$ in order to yield a statement about $\psi$:
A priori, the statement $(\ref{price})$ is in fact  completely compatible with
\begin{equation}
\label{notstable}
\limsup_{t\to\infty} \psi(r,t)=\infty.
\end{equation}

\section{Kerr}
Now that we have a decay result for Schwarzschild, can we go back and
retrieve this for Kerr?

Unfortunately, like the boundedness proof, 
our decay proof too is unstable, but for a different reason: \emph{The structure
of trapping in Schwarzschild is very special.}
In particular, the construction of $X$ in Section~\ref{Xconstr}
is based on the fact that the co-dimensionality of the set of trapped
null geodesics manifests itself also in physical space in the following way: all such 
trapped geodesics approach the codimension-one hypersurface $r=3M$.
(Recall that the function $f_b$ of $(\ref{fdef})$ vanishes precisely along this hypersurface.)
See also~\cite{alinhac} for a nice discussion of this issue.

Nonetheless, it turns out that using ideas from the decay proof,
we can indeed perturb \emph{just the boundedness theorem} 
for geometries $g$ ``near'' Schwarzschild, provided
that $g$ retains two of the Killing fields of Schwarzschild ($T$ and $\Omega_1$ say), and a certain
geometric property. 

Unlike the theorem of Section~\ref{perturb?}, the
class of spacetimes allowed will in particular include the Kerr case for $a\ll M$.

\subsection{Uniform boundedness on axisymmetric stationary black hole
exteriors}
\label{Ubound}
Let us first state the theorem
\begin{theorem*}~\cite{bounded}
Let  $\mathcal{R}$ be as before, $g$ be a metric defined on 
$\mathcal{R}$, and let
$T$ and $\Phi=\Omega_1$ be Schwarzschild Killing fields.
Assume
\begin{enumerate}
\item
$g$ is close to Schwarzschild in an appropriate sense
\item
$T$ and $\Phi$ are Killing with respect to $g$
\item
$\mathcal{H}^+$ is null with respect to $g$, and
$T$ and $\Phi$ together span the null generator of $\mathcal{H}^+$.
\end{enumerate}
Then the uniform boundedness theorem of Section~\ref{stronger} holds.
\end{theorem*}
In particular, the theorem applies to Kerr for $|a|\ll M$, Kerr-Newman for
$|a|\ll M$, $Q\ll M$, etc.

The heuristic idea of the proof of this result is actually quite simple.
Consider a metric $g$ as described above, i.e.~retaining the Killing fields $T$
and $\Phi$ of Schwarzschild, and suitable close to Schwarzschild.

Via the Fourier transfrom, we associate frequencies
$\omega$, $k$ to the Killing fields $T$ and $\Phi$, 
where $\omega\in \mathbb R$, and $k\in \mathbb Z$. 
Suppose we
 could decompose 
 \begin{equation}
 \label{decompose}
 \psi=\psi_\sharp+\psi_\flat
 \end{equation}
 where $\hat\psi_\flat$ is supported in $\omega^2\le c k^2$
 and $\hat\psi_\sharp$ is supported in $\omega^2\ge c k^2$.

 The crucial observation is simply the following:
\emph{For $c$ small enough, and for $g$ close enough to 
Schwarzschild, then in view of the geometric assumption 3.~on the Killing fields,
it follows
that (i)
there is no superradiance for $\psi_\sharp$, and (ii)
there is no trapping for  $\psi_\flat$.}

That is to say, for appropriate choice of $c$,
(i) the current
$J^T_\mu(\psi_\sharp)$ has  a nonnegative flux through the horizon $\mathcal{H}$,
and (ii) 
a variant of the $X$ vector field can be constructed, so that $K^X(\psi_\flat)$
is nonnegative. In view of the absence of trapping, the current $K^X(\psi_\flat)$ need
not degenerate near $r=3M$, and its construction is quite simple relative to Section~\ref{Xconstr}, 
and moreover, completely stable to perturbation. In particular, it suffices to know that
such a current can be constructed on Schwarzschild giving the required positivity
properties in this frequency range.

Thus, the outline of the boundedness argument appears quite simple:
Apply $T$ and $N$ to $\psi_\sharp$ as in the boundedness proof, and apply
$T$, $N$, and $X$ to $\psi_\flat$ as in the decay proof to obtain 
integrated decay (and thus in particular energy boundedness!)
for $\psi_\flat$.  This would in particular yield the non-degenerate energy boundedness statement
for $\psi=\psi_\sharp+\psi_\flat$. The pointwise estimates would then follow
by commutation, in view also of Section~\ref{red}.

To implement the above argument, however, is tricky: In order to decompose $\psi$ as
in $(\ref{decompose})$ one would in particular have to take the Fourier transform of $\psi$ in time.
Yet a priori we have not  shown that $\psi$ is even uniformly bounded. 
Thus we must replace $\psi$ with a cut-off version $\psi_{\hbox{\Rightscissors}}=  \xi  \psi$,
 where
$\xi$ is a cutoff function in time, and apply the decomposition to
$\psi_{\hbox{\Rightscissors}}$. generating error terms which must
themselves be bounded.  It is essential that one has at ones disposal a non-degenerate
energy, as in the statement of theorem, to bound these error terms.
This is accomplished via a bootstrap argument. See~\cite{bounded}
for the details of the proof.

\subsection{Decay for slowly rotating Kerr}
\label{DecKer}
The above argument for boundedness
is relatively simple and robust because it circumvents the problem 
of understanding trapping: It sufficed to know that $\psi_\flat$ is not trapped.
If one is to tackle the problem of decay, however,
one has no choice but to come to terms with the structure of trapping in detail.
Since the codimensionality of the trapping must be viewed in phase space, 
this suggests adapting our arguments, particularly the construction of $X$,
to phase space. We shall be able to accomplish this, but at the expense
of restricting to Kerr spacetimes, as opposed to the general class of Section~\ref{Ubound}.

\subsubsection{The separation and the frequency-localised construction of $X$}
\label{separ}
There is a convenient way of doing phase space analysis in Kerr spacetimes, namely,
as discovered by Carter~\cite{cartersep2}, 
the wave equation can be separated.  Walker and Penrose~\cite{walker} later showed that
both the complete integrability of geodesic flow and the separability of the wave equation
have their fundamental origin in the presence of a \emph{Killing tensor}.
In fact,  as we shall see, in view of its intimate relation with the 
integrability of geodesic flow, Carter's separation of $\Box_g$ immediately captures
the codimensionality of the trapped set.

The separation of the wave equation requires taking the Fourier transform with respect to
time, and
then expanding into oblate spheroidal harmonics. 
As before, taking the Fourier transform requires cutting off in time.
Since this has essentially already been addressed in the previous section, let us pretend that
this is not an issue, and that we may write
\[
\hat\psi(\omega,\cdot)=
\sum_{m,\ell}R^\omega_{m\ell}(r) S_{m\ell}(a\omega,\cos\theta)e^{im\phi^*},
\]
where $S_{m\ell}$ are the oblate spheroidal harmonics with eigenvalues $\lambda_{m\ell}(\omega)$.
The wave equation $(\ref{thewaveequation})$ then reduces
 to the following equation for $R_{m\ell}^\omega$:
\[
\Delta \frac{d}{dr} \left (\Delta \frac{R_{m\ell}^\omega}{dr}\right) + \left (a^2m^2 + (r^2+a^2)^2\omega^2-\Delta(\lambda_{m\ell}+a^2\omega^2) \right) R_{m\ell}^\omega=0.
\]

Defining a coordinate $r^*$ by
$\frac{dr^*}{dr}=\frac{r^2+a^2}{\Delta}$,
where $\Delta=r^2-2Mr+a^2$,
and setting
$u(r)=(r^2+a^2)^{1/2}
 R^\omega_{m\ell} (r)$,
then $u$ satisfies
\[
\frac{d^2}{(dr^*)^2}u+(\omega^2 - V^\omega_{m\ell }(r))u = 0
\]
where
\[
V^\omega_{m \ell}(r)= \frac{4Mram\omega-a^2m^2+\Delta (\lambda_{m\ell}+\omega^2a^2)}{(r^2+a^2)^2}
+\frac{\Delta(3r^2-4Mr+a^2)}{(r^2+a^2)^3}
-\frac{3\Delta^2 r^2}{(r^2+a^2)^4}.
\]
Consider the following quantity 
$$
Q=
f \left ( \left|\frac {du}{dr^*}\right|^2 + (\omega^2-V ) |u|^2\right) + \frac {df}{dr^*} 
{\rm Re}\left(\frac{du}{dr^*} \bar u\right)-
\frac 12 \frac{d^2f}{{dr^*}^2} |u|^2.
$$
Then, with the notation $'=\frac{d}{dr^*}$,
\begin{equation}\label{eq:Q}
Q'= 2 f' |u'|^2  - f V' |u|^2 -\frac 12 f''' |u|^2.
\end{equation}

The main difficulty is for $\hat\psi$
supported in $|\omega|\ge \omega_1$, 
$\lambda_{m\ell} \ge \lambda_2\omega^2$,
where we may choose $\omega_1$ (but not $\lambda_2$) large.
An easy computation shows that for suitable choice of $\omega_1$, in this frequency
range $V'$ has a unique simple zero. Let us denote the $r$-value of this zero by
$r_{m\ell}^\omega$.

We now choose $f$ so that (i)
$f'\ge 0$, (ii) $f\le 0$ for $r\le r_{m\ell}^\omega$ and $f\ge 0$ for $r\ge r_{m\ell}^\omega$ ,
and (iii)
 $-fV'-\frac 12 f'''\ge c>0$.

Integrating the identity \eqref{eq:Q} and using that $u\to 0$ as $r\to \infty$ we obtain that for any compact
set $K_1$ in $r^*$ and a certain compact set $K_2$ (which in particular does not contain
$r=3M$), there exists a positive constant
$b>0$ so that
\begin{align*}
b\int_{K_1} &(|u'|^2+|u|^2) dr + b(\lambda_{m\ell}+\omega^2) 
\int_{K_2} |u|^2 dr
\le \left (|u'|^2+(\omega^2-V) |u|^2\right)(r_+),
\end{align*}
where $r_+$ denotes the $r$-value of $\mathcal{H}^+$.
Reinstating the dropped indices $m,\ell,\omega$, suming over $m$, $\ell$,
integrating over $\omega$, and adding this estimate to an estimate for the 
remaining frequencies (which in fact need not degenerate near $r=3M$),
and finally adding a little bit of
the estimate corresponding to $N$ (recall that the computation $(\ref{upologismos})$ is stable!),
we obtain the analogue of $(\ref{finalestimate})$ 
for $\psi$ (with $T\psi$ replacing $\Omega_i\psi$).

This yields integrated decay for solutions to $(\ref{thewaveequation})$ on Kerr $g_{a,M}$
with small $a\ll M$.

\subsubsection{The use of $N$ and $Z$}
Once one has the integrated decay estimates, the other aspects of the proof
of decay on Schwarzschild are stable to perturbation of the metric, 
modulo a loss in  $\delta$ in the $\tau$ power of
the rate of decay, where $\delta$ depends on the closeness to Schwarzschild.
This requires, however, a refinement of the use of $N$, in view of the fact that
the vector field $Z$ as defined in $(\ref{Zdef})$ fails to be $C^1$ on
$\mathcal{H}^+$. A further issue arises in that one must commute with
the Schwarzschild $\Omega_i$ to obtain the desired pointwise decay statements,
and these are no longer Killing, generating errors.
See~\cite{notes} for details.

\subsubsection{The statement of the theorem}
\label{statthe}
\begin{theorem*}~\cite{notes} 
\label{DT}
Let $(\mathcal{M},g_{a,M})$ be Kerr for $|a|\ll M$, $\mathcal{D}$ be the closure of its
domain of dependence,
let $\Sigma_0$ be the surface $\mathcal{D}\cap\{t^*=0\}$, 
let $\upPsi$, $\upPsi'$ be initial data on $\Sigma_0$ such that
$\upPsi\in H^s_{\rm loc}(\Sigma)$, $\upPsi'\in H^{s-1}_{\rm loc}(\Sigma)$ for $s\ge 1$,
and $\lim_{x\to i^0}\upPsi=0$,
and let $\psi$ be the 
corresponding unique solution
of $\Box_g\psi=0$. Let $\varphi_\tau$ denote the $1$-parameter
family of diffeomorphisms generated by $T$,
let $\tilde\Sigma_0$ be a spacelike hypersurface in $J^+(\Sigma_0)$
terminating on null infinity,
and define $\tilde\Sigma_\tau=\varphi_\tau (\tilde\Sigma_0)$.
Let $s\ge 3$ and assume
\[
E_1\doteq 
\int_{\Sigma_0} r^2 (J_\mu^{n_0}(\psi) + J_\mu^{n_0}(T\psi)+ J_\mu^{n_0}(TT\psi)) n^\mu_0<\infty.
\]
Then there exists a $\delta>0$ depending on $a$ (with $\delta\to 0$ as $a\to 0$) and a
$B$ depending only on $\tilde\Sigma_0$ such that
\[
\int_{\tilde \Sigma_\tau} J^N_\mu (\psi) n^\mu_{\tilde{\Sigma_\tau}} \le
BE_1\, \tau^{-2+2\delta}.
\]
Now let $s\ge 5$ and
assume
\[
E_2\doteq\sum_{|\alpha|\le2 } \sum_{\Gamma=\{T, N, \Omega_i\}}
\int_{\Sigma_0} r^2 (J_\mu^{n_0}(\Gamma^\alpha \psi) + J_\mu^{n_0}(\Gamma^\alpha T\psi)+ J_\mu^{n_0}(\Gamma^\alpha TT\psi)) n^\mu_0
<\infty
\]
where $\Omega_i$ are the angular momentum operators corresponding to the related
Schwarzschild metric $g_M$ on $\mathcal{D}$.
Then
\[
\sup_{\tilde\Sigma_\tau}\sqrt{r}|\psi|\le B \sqrt{E_2}\, \tau^{-1+\delta}, \qquad
\sup_{\tilde\Sigma_\tau}r|\psi|\le B \sqrt{E_2}\, \tau^{(-1+\delta)/2}.
\]
\end{theorem*}
One can obtain decay for arbitrary derivatives, including 
transversal derivatives to $\mathcal{H}^+$,
using additional commutation by $N$. 


The above theorem was first announced at the Clay Summer School in Z\"urich in July 2008
and its proof appears in the lecture notes~\cite{notes}. 

There is some additional interesting work in progress related to this section which should be noted:
Tohaneanu et al. are pursuing a related approach to the integrated
decay statement of Section~\ref{separ}, 
again relying on the red-shift estimates developed in~\cite{dr3, bounded}
(presented here in Section~\ref{red}),
but where the frequency localisation is carried out with the machinery of the 
pseudodifferential calculus.\footnote{Private communication from M.~Tohaneanu who
attended~\cite{notes}.}
Andersson and Blue are pursuing an alternative approach
to  the construction of Section~\ref{separ},
in terms of higher order currents, similar to  the current $(\ref{thecurrent})$ of 
Section~\ref{Xconstr}, 
but where
commutation with $\Omega_i$ is repalced by commutation with the so-called Carter
operator.\footnote{Lecture of P.~Blue, Stockholm, September 2009.}

\section{Future directions and the non-linear stability of the Kerr family}
\label{last}
The theorem of Section~\ref{statthe} does not close the book on this subject. 
It would be nice to obtain this result with the least possible assumptions on the geometry.
For instance, what can be said about decay under the much more general
assumptions of our boundedness result, the theorem of Section~\ref{Ubound}? 
It would be interesting also to obtain stronger decay rates in the interior.
In the Kerr case, it is important to obtain results for the whole range 
$a< M$.\footnote{While
no statement is known for general solutions in this range, the following pretty
non-quantitative statement away from the horizon
is given in~\cite{fksy, fksy2} for individual azimuthal modes
$\psi_m$: \emph{If  $\psi_m$ is not supported in a neighborhood of $i^0$ and
$\mathcal{H}^+\cap\mathcal{H}^-$, then for $r>r_+$, $\lim_{t\to\infty} \psi_m(r,t)=0$.}  
As in the heuristics of Price, this statement is of course
compatible with~$(\ref{notstable})$ for the sum over $m$.}
Moreover, it is essential to understand boundedness and decay properties for
higher spin (see below).
Another interesting direction is to study spacetimes with cosmological constant
(see~\cite{kostakis}). For an extensive list 
of related open problems, see~\cite{notes}.

The most important future direction, however, and the main motivation for the
problem considered in  this talk is
the stability of the Kerr family of spacetimes
as solutions to the Cauchy problem for the Einstein vacuum equations
\[
R_{\mu\nu}=0.
\]
See~\cite{notes} for a formulation. 
This latter problem is one of the main open problems in 
general relativity.

The role of linear theory for the understanding of the non-linear stability problem
can be seen from 
the proof of the  nonlinear stability of Minkowski space, first given in
Christodoulou--Klainerman~\cite{book}.
The proof of~\cite{book}
required in particular a robust method of proving
the results of Sections~\ref{Ubound} 
and~\ref{DecKer}, 
not just for the wave
equation $(\ref{thewaveequation})$, 
but for the spin-$2$ Bianchi system satisfied by the curvature tensor, 
and not just on a background which was exactly Minkowski,
but for spacetimes sufficiently close to and decaying to Minkowski. 
Hence the importance
of the open problems discussed above.

\begin{theacknowledgments}
The material presented here is in collaboration with Igor Rodnianski.
Some of the text has been adapted from our~\cite{notes}.
\end{theacknowledgments}


\begin{thebibliography}{9}
\bibitem{alinhac} S. Alinhac \emph{Energy multipliers for perturbations of Schwarzschild metric}
preprint, 2008

\bibitem{bachelot} A. Bachelot \emph{Asymptotic completeness for the Klein-Gordon
equation on the Schwarzschild metric}, Ann. Inst. H. Poincar\'e Phys. Th\'eor. {\bf 16}
(1994), no. 4, 411--441

\bibitem{BlueSof0} P. Blue and A. Soffer  \emph{Semilinear wave
equations on the Schwarzschild manifold. I. Local decay
estimates}, Adv. Differential Equations {\bf 8} (2003), no. 5, 595--614

\bibitem{BlueSof} P. Blue and A. Soffer \emph{Errata for ``Global existence \ldots Regge Wheeler equation''}, gr-qc/0608073

\bibitem{BlueSter} P. Blue and J. Sterbenz  \emph{Uniform decay
of local energy and the semi-linear wave equation
on Schwarzschild space} Comm. Math. Phys. {\bf 268} (2006), no. 2,
481--504


\bibitem{cartersep2}
B. Carter \emph{Hamilton-Jacobi and Schr\"odinger separable solutions of Einstein's
equations} Comm. Math. Phys. {\bf 10} (1968), 280--310


\bibitem{book2} D. Christodoulou \emph{The action principle and partial
differential equations}, Ann. Math. Studies No. 146, 1999 

\bibitem{book}  
D. Christodoulou and S. Klainerman \emph{The global nonlinear
stability of the Minkowski space} Princeton University Press, 1993


\bibitem{dr1} M. Dafermos and I. Rodnianski
\emph{A proof of Price's law for the collapse of a 
self-gravitating scalar field}, Invent. Math. {\bf 162}
(2005), 381--457


\bibitem{dr3} M. Dafermos and I. Rodnianski
\emph{The redshift effect and radiation decay on black hole
spacetimes}, gr-qc/0512119

\bibitem{dr4} M. Dafermos and I. Rodnianski
\emph{The wave equation on  Schwarzschild-de Sitter spacetimes},
arXiv:0709.2766v1 [gr-qc]

\bibitem{dr5} M. Dafermos and I. Rodnianski
\emph{A note on energy currents and decay for the wave equation
on a Schwarzschild background}, arXiv:0710.0171v1 [math.AP]

\bibitem{bounded} M. Dafermos, and I. Rodnianski, \emph{A proof of the uniform 
boundedness of solutions to the wave equation on slowly
rotating Kerr backgrounds},
available online at \url{http://arxiv.org/abs/0805.4309}

\bibitem{notes} M. Dafermos, and I. Rodnianski, \emph{Lectures on black holes
and linear waves}, to appear in Clay Lecture Notes,
available online at \url{http://arxiv.org/abs/0811.0354}


\bibitem{fksy} F. Finster, N. Kamran, J. Smoller, 
S. T. Yau \emph{Decay of solutions of the wave equation in Kerr geometry}
Comm. Math. Phys. {\bf 264} (2006), 465--503

\bibitem{fksy2} F. Finster, N. Kamran, J. Smoller, S.-T. Yau
\emph{Erratum: Decay of solutions of the wave equation in Kerr geometry}
Comm. Math. Phys., online first


\bibitem{he:lssst}
S. W. Hawking and G. F. R. Ellis
\emph{The large scale structure of space-time} Cambridge
Monographs on Mathematical Physics, No. 1. Cambridge
University Press, London-New York, 1973

\bibitem{kostakis}
G. Holzegel \emph{On the massive wave equation on slowly rotating
Kerr-AdS spacetimes}, preprint 2009

\bibitem{kw:lss} B. Kay and R. Wald
\emph{Linear stability of Schwarzschild under perturbations
which are nonvanishing on the bifurcation $2$-sphere}
Classical Quantum Gravity {\bf 4} (1987), no. 4, 893--898

\bibitem{muchT} S. Klainerman \emph{Uniform decay estimates and the Lorentz
invariance of the classical wave equation} Comm. Pure Appl. Math. {\bf 38} (1985),
321--332

\bibitem{labasoffer} I. Laba and A. Soffer \emph{Global existence and scattering
for the nonlinear Schr\"odinger equation on Schwarzschild manifolds}
Helv. Phys. Acta {\bf 72} (1999), no. 4, 272--294

\bibitem{metcalfe} J. Metcalfe \emph{Strichartz estimates on Schwarzschild space-times
(joint work with D.~Tataru, M.~Tohaneanu)}
Oberwolfach Reports {\bf 44} (2007),  8--11.

\bibitem{os} J. R. Oppenheimer and H. Snyder \emph{On continued gravitational
contraction} Phys. Rev. {\bf 56} (1939), 455--459

\bibitem{rpr:ns} R. Price \emph{Nonspherical perturbations
of relativistic gravitational collapse. I. Scalar and gravitational
perturbations} Phys. Rev. D (3) {\bf 5} (1972), 2419--2438

\bibitem{RW} T. Regge and J. Wheeler \emph{Stability of a Schwarzschild singularity}
Phys. Rev. {\bf 108} (1957), 1063--1069

\bibitem{twainy} F. 
Twainy \emph{The Time Decay of Solutions to the Scalar Wave Equation in Schwarzschild
Background}
Thesis. San Diego: University of California 1989

\bibitem{wald} R. Wald \emph{General relativity}
University of Chicago Press, Chicago, 1984

\bibitem{walker}
M. Walker and R. Penrose 
\emph{On quadratic first integrals of the geodesic equations for type ${22}$ spacetimes} 
Comm. Math. Phys. {\bf 18} (1970), 265--274


\bibitem{whiting} B. Whiting \emph{
Mode stability of the Kerr black hole}
J. Math. Phys. {\bf 30} (1989), 1301

\end{thebibliography}
\end{document}